\documentclass[prb,aps,twocolumn,amsmath,amssymb,floatfix,showpacs,
superscriptaddress]{revtex4}

\usepackage[dvips]{graphics}
\usepackage{color}
\definecolor{dred}{rgb}{0,0,0.6}

\begin{document}

\title{\textcolor{dred}{Conformation-dependent electron transport through 
a biphenyl molecule: Circular current and related issues}}

\author{Santanu K. Maiti}

\email{santanu.maiti@isical.ac.in}

\affiliation{Physics and Applied Mathematics Unit, Indian Statistical
Institute, 203 Barrackpore Trunk Road, Kolkata-700 108, India}

\begin{abstract}

We investigate the conformation-dependent electron transfer in a biphenyl 
molecule within a simple tight-binding framework. The overall junction 
current and circular currents in two benzene rings driven by applied
bias voltage are calculated by using Green's function formalism. Our
analysis may provide the possibilities of using organic molecules with 
loop substructures to design molecular spintronic devices, indicating
the emergence of molecular spintronics. 

\end{abstract}


\maketitle

\section{Introduction}

The study of electron transport through a molecular system attached to
metallic electrodes is regarded as one of the most promising research fields 
in nanoscale technology and physics~\cite{nitzan1}. With the discovery of 
advanced molecular scale measurement methodologies like scanning 
electro-chemical microscopy (SESM), scanning tunneling microscopy (STM), 
atomic force microscopy (AFM), etc., it is now possible to measure current 
flow through single molecules or cluster of molecules sandwiched between two 
electrodes~\cite{chen}. The proposed idea of designing molecule-based diode 
by Aviram and Ratner~\cite{aviram} in 1974 first illustrates the possibility 
of using molecules as active components of a device. Since then several 
ab-initio as well as model calculations have been done to investigate
molecular transport theoretically~\cite{ventra1,ventra2,sumit,tagami,orella1,
orella2,arai,walc,san1,san2,san3,san4}. But experimental realizations took 
a little longer time to get feasible. In a pioneering experiment, Reed 
{\em et al.} investigated current-voltage ($I$-$V$) characteristics in 
a single benzene molecule coupled to metallic electrodes via thiol 
groups~\cite{reed2}. 
Various other experiments using molecules have also been reported in 
literature exploring many interesting features e.g., ballistic transport, 
quantized conductance, negative differential resistance (NDR), gate 
controlled transistor operation, memory effects, bistable switching, 
conformational switching to name a few. Although a lot of 
theoretical~\cite{mag,lau,baer1,baer2,baer3,gold} and 
experimental~\cite{reed1,tali,fish,cui,gim} studies have been made so far 
using different molecules, yet several problems are to be solved for  
further development of molecular electronics.

Most of the works associated with electronic conduction through molecular 
bridge systems are mainly concerned on the overall conduction properties. 
But a very few works are available where attention has been paid to the 
current distribution within the molecule itself having single or multiple 
loop substructures~\cite{dist1,dist2,dist3}. Recently some interesting 
works have been done by Nitzan {\em et al.} and few other groups where 
possible quantum interference effects have been explored on current 
distribution through such molecular geometries due to the existence of 
multiple pathways, yielding the possibilities of voltage driven circular 
currents~\cite{cir1,cir2,cir3,cir4,cir5,cir6}. The appearance of circular
currents in loop geometries have already been reported in other contexts
several years ago. This is commonly known as persistent currents in 
mesoscopic conducting rings where the current is induced by means of 
Aharonov-Bohm flux $\phi$ threaded by the ring~\cite{butt2,levy,gefen,skm1,
skm2,skm3,skm4,skm5,skm6}. The reason behind this current in isolated loop 
geometries is 
quite different from the previous one where current is driven by an applied 
bias voltage. It has been verified that the circular currents appear in 
molecular rings, driven by applied bias voltage, produce considerable 
magnetic fields at the centers of these rings. This phenomenon is somewhat 
interesting and can be used in different aspects in the study of molecular 
transport. For example, in presence of a local spin at the ring center one 
can regulate spin dependent transport through the molecular wire by tuning 
the orientation of that local spin, and, also the behavior of spin inelastic 
currents can be explained. In a recent work Galperin {\em et al.}~\cite{rai} 
have proposed some results towards this direction. One can also utilize this 
circular current generated magnetic field in other way to control spin 
dependent transport through a molecular wire without changing the orientation 
of the local spin. In that case we can change the strength of the magnetic 
field by some mechanisms. To test it, biphenyl molecule may the best 
example where two benzene rings are connected by a single C-C bond. It 
has been examined~\cite{latha} that in the case of a biphenyl molecule, 
electronic conductance changes significantly with the relative twist angle 
among two benzene rings. For the planar conformation
conductance becomes maximum, while it gets decreased with increasing the
twist angle. This phenomenon motivates us to describe conformation-dependent
circular currents in a biphenyl molecule coupled to two metallic electrodes.
We use a simple tight-binding (TB) framework to describe the model quantum
system and evaluate all the results through Green's function formalism.
We believe that our present analysis will certainly provide some important 
information that can be used to design molecular spintronic devices in 
near future. 

The structure of the paper is as follows. In section II, we describe the
molecular model and theoretical formulation for the calculations. The 
essential results are presented in Section III which contains (a) 
transmission probability as a function of injecting electron energy and 
junction current through the molecular wire as a function of applied
bias voltage for different twist angles, and (b) conformation-dependent
circular currents in two benzene rings and associated magnetic fields 
at the ring centers. Finally, in section IV, we summarize our main results 
and discuss their possible implications for further study.

\section{Molecular Model and Theoretical Formulation}

\subsection{Tight-binding model}

\begin{figure}[ht]
{\centering \resizebox*{7.5cm}{2.2cm}{\includegraphics{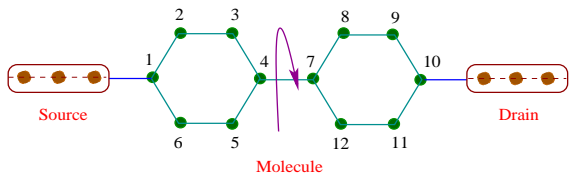}}\par}
\caption{(Color online). Schematic diagram of a biphenyl molecule attached
to two electrodes, namely, source and drain. The magenta arrow describes 
the relative twist among the molecular rings.}
\label{biphenyl}
\end{figure}

Figure~\ref{biphenyl} gives a schematic illustration of the molecular wire, 
where a biphenyl molecule is coupled to two semi-infinite one-dimensional 
($1$D) metallic electrodes, commonly known as source and drain. Our analysis 
for the present work is based on non-interacting electron picture, and, 
within this framework, TB model is extremely suitable for analyzing electron
transport through a molecular bridge system. 
The single particle Hamiltonian which captures the molecule and side-attached
electrodes gets the form:
\begin{equation}
H=H_{mol} + H_{ele} + H_{tun}.
\label{equ1}
\end{equation}
The first term $H_{mol}$ corresponds to the Hamiltonian of the biphenyl 
molecule sandwiched between two electrodes. Under nearest-neighbor hopping 
approximation, the TB Hamiltonian of the molecule composed of $12$ ($N=12$) 
atomic sites reads,
\begin{eqnarray}
H_{mol} & = & \sum_i \epsilon c_i^{\dagger} c_i + \sum_i v 
\left[c_{i+1}^{\dagger} c_i + c_i^{\dagger} c_{i+1}\right] \nonumber \\
 & + & \sum_j \epsilon c_j^{\dagger} c_j + \sum_j v
\left[c_{j+1}^{\dagger} c_j + c_j^{\dagger} c_{j+1}\right] \nonumber \\
 & + & v_{4,7} \left[c_4^{\dagger}c_7 + c_7^{\dagger}c_4\right]
\label{equ2}
\end{eqnarray}
where the indices $i$ and $j$ are used for the left and right molecular
rings, respectively. $\epsilon$ denotes the on-site energy of an electron 
at $i$-($j$-)th site and $v$ describes the isotropic nearest-neighbor 
coupling between the molecular sites. $c_i^{\dagger}$($c_j^{\dagger}$) 
and $c_i$($c_j$) are the creation and annihilation operators, respectively, 
of an electron at the $i$-($j$-)th site. The last term in the right hand 
side of Eq.~\ref{equ2} describes the coupling between two molecular rings.
In terms of the relative twist angle $\theta$ among these two rings,
the coupling strength $v_{4,7}$ gets the form: $v_{4,7}=v \cos\theta$.

The second and third terms of Eq.~\ref{equ1} describe the Hamiltonians 
for the $1$D semi-infinite electrodes (source and drain) and 
molecule-to-electrode coupling. In Wannier basis representation they are
expressed as follows.
\begin{eqnarray}
H_{ele} & = & H_S + H_D \nonumber \\
 & = & \sum_{\alpha=S,D} \left\{\sum_n \epsilon_0 d_n^{\dagger} d_n 
+ \sum_n t_0 \left[d_{n+1}^{\dagger} d_n + h.c. \right]\right\}, 
\nonumber \\
\label{equ3}
\end{eqnarray}
and,
\begin{eqnarray}
H_{tun} & = & H_{S,mol} + H_{D,mol} \nonumber \\
& = &   \tau_S[c_p^{\dag}d_0 + h.c.] + \tau_D[c_q^{\dag}d_{N+1} + h.c.].
\label{equ4}
\end{eqnarray}
Here, $\epsilon_0$ and $t_0$ correspond to the site energy and 
nearest-neighbor hopping integral in the electrodes. $d_{n}^{\dag}$ and 
$d_{n}$ are the creation and annihilation operators, respectively, of an 
electron at the site $n$ of the electrodes. The coupling strength between 
the source and the molecule is $\tau_S$, while it is $\tau_D$ between the 
molecule and the drain. The source and drain are coupled to the molecule 
through $p$-th and $q$-th atomic sites, respectively, those are variable.

\subsection{Two-terminal transmission probability and junction current}

To obtain transmission probability of an electron from source to drain
electrode through the molecule, we use Green's function formalism. Within 
the regime of coherent transport and in the absence of Coulomb interaction
this technique is well applied.

The single particle Green's function operator representing the entire
system for an electron with energy $E$ is defined as,
\begin{equation}
G=\left(E - H + i\eta \right)^{-1}
\label{eqn8}
\end{equation}
where, $\eta \rightarrow 0^+$.
Following the matrix forms of \mbox{\boldmath $H$} and \mbox{\boldmath $G$},
the problem of finding \mbox{\boldmath $G$} in the full Hilbert space of
\mbox{\boldmath $H$} can be mapped exactly to a Green's function
\mbox{\boldmath $\mathcal G$} corresponding to an effective
Hamiltonian in the reduced Hilbert space of the molecule itself and
we have~\cite{datta},
\begin{equation}
\mbox{\boldmath ${\mathcal G}$}=\left(\mbox {\boldmath $E-H_{mol}-\Sigma_S
-\Sigma_D$}\right)^{-1}.
\label{equ9}
\end{equation}
Here, \mbox{\boldmath $\Sigma_S$} and \mbox{\boldmath $\Sigma_D$} are the 
contact self-energies introduced to incorporate the effect of coupling of 
the molecule to the source and drain, respectively. In terms of this 
effective Green's function \mbox{\boldmath ${\mathcal G}$}, two-terminal
transmission probability $T$ through the molecular wire can be written 
as~\cite{datta},
\begin{equation}
T = \mbox{Tr}\mbox{\boldmath [$\Gamma_S \mathcal {G}^r \Gamma_D 
\mathcal {G}^a$]},
\label{eqn13}
\end{equation}
where, \mbox{\boldmath $\Gamma_S$} and \mbox{\boldmath $\Gamma_D$} are the 
coupling matrices, and, \mbox{\boldmath ${\mathcal G}^r$} and
\mbox{\boldmath ${\mathcal G}^a$} are the retarded and advanced Green's
functions, respectively. 

With the knowledge of the transmission probability we compute overall
junction current ($I_T$) as a function of bias voltage ($V$) using the 
standard formalism based on quantum scattering theory.
\begin{equation}
I_T(V) = \frac{2 e}{h} \int \limits_{- \infty}^{\infty} 
T\,[f_S(E)-f_D(E)]\,dE.
\label{eqn20}
\end{equation}
Here, $f_S$ and $f_D$ are the Fermi functions of the source and drain,
respectively. At absolute zero temperature the above equation boils down 
to the following expression.
\begin{equation}
I_T(V) = \frac{2 e}{h} \int \limits_{E_F - \frac{eV}{2}}^{E_F + 
\frac{eV}{2}} T(E) \,dE,
\label{eqn21}
\end{equation}
where, $E_F$ is the equilibrium Fermi energy.

\subsection{Circular current and associated magnetic field}

\begin{figure}[ht]
{\centering \resizebox*{6cm}{4cm}{\includegraphics{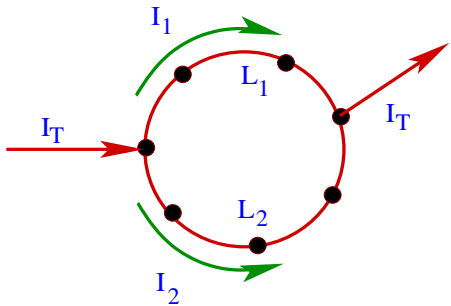}}\par}
\caption{(Color online). Schematic view of current distribution through 
a ring geometry coupled to two electrodes. The filled black circles 
correspond to the positions of the atomic sites.}
\label{ring}
\end{figure}

In order to calculate circular current in molecular rings of the biphenyl 
molecule let us first concentrate on the current distribution in a simple 
loop geometry illustrated in Fig.~\ref{ring}. A net current $I_T$ 
flows between two electrodes through a quantum ring, where $I_1$ and 
$I_2$ are the currents flowing through upper and lower arms of the ring, 
respectively. We assign positive sign to the current propagating in the 
counter-clockwise direction. With this current distribution, we define
circular current of the ring as~\cite{cir1},
\begin{equation}
I_c=\frac{1}{L} \left(I_1 L_1 + I_2 L_2\right)
\label{circur}
\end{equation}
where, $L_1$ and $L_2$ are the lengths of the upper and lower arms of the
ring, respectively, and $L=L_1+L_2$. Thus, in order to compute $I_c$, 
following the above relation, we need to know the currents in different 
branches of the loop geometry. We evaluate these currents by using Green's 
function formalism. At absolute zero temperature the current $I_{ij}$ 
flowing from site $i$ to $j$ ($j=i \pm 1$) is given by the expression.
\begin{equation}
I_{ij}= \int \limits_{E_F - \frac{eV}{2}}^{E_F + 
\frac{eV}{2}} J_{ij}(E) \,dE,
\label{eeee}
\end{equation}
where, $J_{ij}$ is the current density. In terms of the correlation function 
\mbox{\boldmath $\mathcal{G}^n$} it can be written as~\cite{dist3}, 
\begin{equation}
J_{ij} = \frac{4e}{h}\mbox{Im}\mbox{\boldmath [$H_{ij} \mathcal{G}_{ij}^n$]},
\label{ggggg}
\end{equation}
where, \mbox{\boldmath $\mathcal{G}^n$} $=$ \mbox{\boldmath $\mathcal{G}^r
\Gamma_S \mathcal{G}^a$}. This correlation function is evaluated by setting
the occupation function of the source to unity and that of the drain to zero.

Finally, we determine the local magnetic field at any point $\vec{r}$ inside
the ring, associated with circular current $I_c$, from the Biot-Savart's 
Law~\cite{cir1},
\begin{equation}
\vec{B}(\vec{r})=\sum_{(i,j)} \int \frac{\mu_0}{4\pi} I_{ij} 
\frac{d\vec{r} \times (\vec{r}-\vec{r^{\prime}})}
{|(\vec{r}-\vec{r^{\prime}})|^3},
\label{bio}
\end{equation}
where, $\mu_0$ is the magnetic constant and $\vec{r^{\prime}}$ is the 
position vector of an infinitesimal bond current element 
$I_{ij}d\vec{r^{\prime}}$. Using the above expressions we evaluate circular 
currents and associated magnetic fields in two molecular rings of the 
biphenyl molecule.

Throughout this work, we assume that the entire voltage is dropped across 
the molecule-to-electrode interfaces and this assumption is good enough 
for molecules of smaller size. We also restrict ourselves at absolute zero 
temperature and choose the units where $c=e=h=1$.

\section{Numerical Results and Discussion}

In this section we present numerical results computed for transmission
probability, overall junction current and circular currents in a 
biphenyl molecule under conventional biased conditions. Throughout our
analysis we set the on-site energies in the molecule as well as in the
source and drain electrodes to zero, $\epsilon=\epsilon_0=0$. The
nearest-neighbor coupling strength in the electrodes ($t_0$) is fixed 
at $2$eV, while in the molecule ($v$) it is set at $1$eV. The coupling 
strengths of the molecule to the source and drain electrodes, characterized 
by the parameters $\tau_S$ and $\tau_D$, are also set at $1$eV. We fix the
equilibrium Fermi energy $E_F$ at zero and measure the energy scale in 
unit of $v$.

\subsection{Transmission probability and junction current}

\begin{figure}[ht]
{\centering \resizebox*{7.25cm}{4.5cm}
{\includegraphics{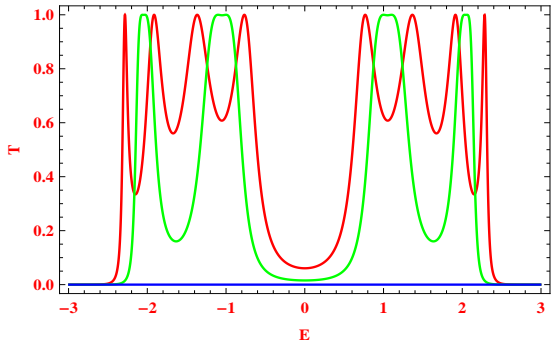}}\par}
\caption{(Color online). Transmission probability as a function of energy
for the biphenyl molecule when the electrodes are connected at the molecular
sites $1$ and $10$, as shown in Fig.~\ref{biphenyl}. The red, green and blue 
lines correspond to $\theta=0$, $\pi/3$ and $\pi/2$, respectively.}
\label{transmission}
\end{figure}

In Fig.~\ref{transmission} we show the variation of transmission 
probability $T$ as a function of injecting electron energy $E$ when the 
source and drain are coupled to the biphenyl molecule at the sites $1$ 
($p=1$) and $10$ ($q=10$), respectively. The red curve corresponds to 
$\theta=0$, while the green and blue lines are associated with $\theta=\pi/3$ 
and $\pi/2$, respectively. 
From the spectrum it is observed that the transmission 
probability exhibits resonant peaks (red and green lines) for some particular 
energies and at these resonances it ($T$) almost reaches to unity. All these 
resonant peaks are associated with the energy eigenvalues of the molecule, 
and therefore, we can say that the transmittance spectrum is a fingerprint 
of the electronic structure of the molecule. The number of
\begin{figure}[ht]
{\centering \resizebox*{7.25cm}{4.5cm}
{\includegraphics{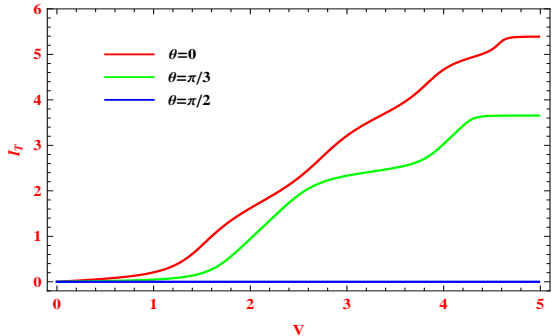}}\par}
\caption{(Color online). Total transport current as a function of bias
voltage when the electrodes are connected to the molecule following the 
same configuration as prescribed in Fig.~\ref{transmission}.}
\label{totalcurrent}
\end{figure}
resonant peaks in $T$-$E$ spectrum and their corresponding 
widths for a particular molecule-to-electrode configuration notably depend 
on the molecular twist angle, which is clearly visible from the red and 
green curves. Depending on the twist angle $\theta$, the resonating energy 
states are available at different energies. For each of these energy 
eigenstates a resonant peak appears in the transmission spectrum with a 
finite width associated with the molecular coupling strength. Now if the 
resonating states with different energies are very closely placed then the 
neighboring peaks can overlap with each other which result a broader peak. 
For low enough molecular coupling, the overlap between neighboring resonant 
peaks is no longer possible, and therefore, separate peaks with identical 
broadening will be obtained in the transmission spectrum. Obviously, for 
different states having identical energy i.e., for degenerate states a 
single resonant peak is generated in the transmission curve. The situation 
especially changes when $\theta=\pi/2$ i.e., one molecular ring becomes 
perpendicular with
\begin{figure}[ht]
{\centering \resizebox*{7.25cm}{4.5cm}
{\includegraphics{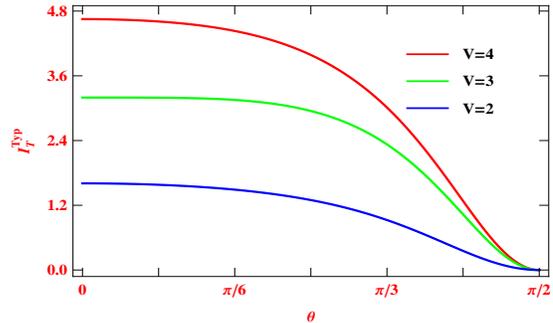}}\par}
\caption{(Color online). Total transport current for a specific voltage
bias as a function of twist angle for the biphenyl molecule. The source
and drain are coupled to the molecular sites $1$ and $10$, respectively.}
\label{typicalcurrent}
\end{figure}
respect to the other ring. In this particular case the transmission 
probability completely disappears for the entire energy band spectrum. 
It is shown by the blue curve in Fig.~\ref{transmission}. With the increment
\begin{figure}[ht]
{\centering \resizebox*{7.25cm}{4.5cm}
{\includegraphics{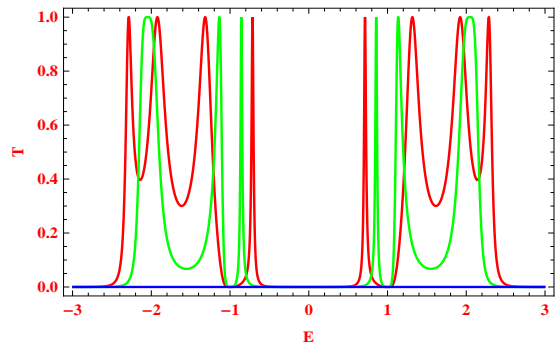}}\par}
\caption{(Color online). Transmission probability as a function of energy
for the same parameter values used in Fig.~\ref{transmission}, when the
source and drain are coupled to the molecular sites $6$ ($p=6$) and $9$
($q=9$), respectively.}
\label{transmission1}
\end{figure}
of the twist angle $\theta$ between these two molecular rings, the degree 
of $\pi$-conjugation between them decreases, which results the reduction
of the junction conductance since the electronic transfer rate through 
the molecule scales as the square of the $\pi$-overlap~\cite{nitz}.
For $\theta=\pi/2$, the $\pi$-conjugation between the molecular rings 
completely disappears, and accordingly, vanishing transmission probability 
is obtained. Thus, rotating one benzene molecule relative to the other one 
can regulate electronic transmission through the biphenyl molecule and 
eventually one can get the insulating phase, which leads to the possibility 
of getting a switching action using this molecule.

The sharpness of the resonant peaks in $T$-$E$ spectrum strongly depends
on the molecular coupling strength to side-attached electrodes, and, it 
greatly controls electron transfer through the bridge system.
In the limit of weak-coupling i.e., $\tau_S(\tau_D)<<v$, sharp resonant
peaks are observed in transmission spectrum, whereas widths of these
peaks get broadened in the limit of strong-coupling, $\tau_S(\tau_D)\sim v$.
\begin{figure}[ht]
{\centering \resizebox*{7.25cm}{4.5cm}
{\includegraphics{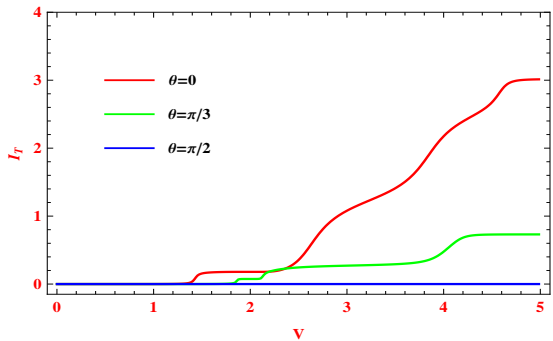}}\par}
\caption{(Color online). Total transport current as a function of bias
voltage when the electrodes are coupled to the molecule following the
same configuration as mentioned in Fig.~\ref{transmission1}.}
\label{totalcurrent1}
\end{figure}
The broadening of transmission peaks with the enhancement in coupling 
strength is quantified by the imaginary parts of the self-energy matrices
{\boldmath $\Sigma_S$} and {\boldmath $\Sigma_D$} which are incorporated 
in the transmittance expression via the coupling matrices 
{\boldmath $\Gamma_S$} and {\boldmath $\Gamma_D$}. This coupling effect 
on electron transport has already been explored elaborately in 
literature~\cite{sm1,sm2,sm3,sm4}, and therefore, in the present work we 
do not illustrate it further.

The fundamental features of electron transport will be more transparent from 
our current-voltage characteristics. The overall junction  current $I_T$
\begin{figure}[ht]
{\centering \resizebox*{8cm}{2.25cm}
{\includegraphics{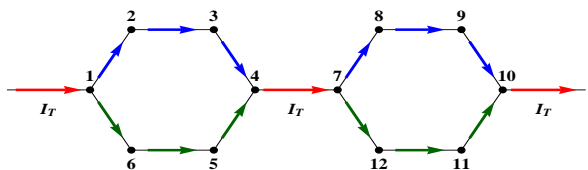}}\par}
\caption{(Color online). Internal current distribution in the molecule
for its planar conformation when the bias voltage is fixed at $4$V. The
source and drain are attached to the atomic sites $1$ and $10$, 
respectively.}
\label{distribution1}
\end{figure}
through the molecular wire following Landauer-B\"{u}ttiker formalism 
(Eq.~\ref{eqn21}). In Fig.~\ref{totalcurrent} we display the variation of 
junction current $I_T$ as a function of applied bias voltage $V$ for the 
biphenyl molecule where the electrodes are connected at the molecular sites 
$1$ and $10$, same as in Fig.~\ref{biphenyl}. The current varies quite 
continuously with the 
voltage bias. Depending on the molecular coupling to the source and drain 
electrodes, the current exhibits continuous-like or step-like behavior 
since it is computed by integrating over the transmission 
curve. For the weak molecular coupling step-like nature associated with
sharp resonant peaks in transmission spectrum will be obtained, unlike to 
the continuous-like feature as observed in the limit of strong-coupling.
Therefore, for a fixed voltage bias one can regulate the current amplitude 
by tuning molecular coupling strength, and, this phenomenon provides an 
interesting behavior in designing molecular electronic devices. 
Figure~\ref{totalcurrent} reveals that the junction current decreases with 
increasing the relative twist angle, following the $T$-$E$ characteristics. 
In addition to this behavior, it is also important to note that the threshold 
bias voltage $V_{th}$ of electron conduction firmly depends on the twist 
angle $\theta$, which is clearly noticed by comparing the red and green 
curves in Fig.~\ref{totalcurrent}.

In order to explore the dependence of electron conduction through the 
biphenyl molecule for any arbitrary angle of twist, in 
\begin{figure}[ht]
{\centering \resizebox*{6cm}{3.5cm}
{\includegraphics{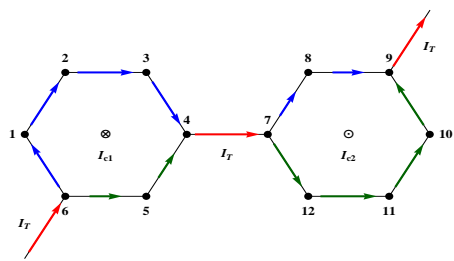}}\par}
\caption{(Color online). Internal current distribution in the molecule for 
its planar conformation when the electrodes are coupled to the atomic sites 
$6$ and $9$, for the same bias voltage taken in Fig.~\ref{distribution1}.
The directions of the corresponding magnetic fields at the ring centers 
are illustrated by the encircled cross and dot representing downward (into
page) and upward (out of page) directions, respectively.}
\label{distribution2}
\end{figure}
Fig.~\ref{typicalcurrent} we present the variation of junction current 
as a function of relative twist angle $\theta$ for some typical bias 
voltages. The spectrum shows that for the planar conformation total 
current amplitude becomes maximum and it gradually decreases with the 
relative twist angle and eventually drops to zero when $\theta=\pi/2$. 
Thus, at $\theta=\pi/2$ no electron conduction will take place through 
this molecular bridge system, while for other choices of $\theta$ electron 
can transfer through the molecule from the source to drain electrode, 
which promotes a conformation-dependent switching action using the 
biphenyl molecule.
  
A significant change in the transmission spectrum is realized when the
electrodes are coupled to the molecule in such a way that the upper and
lower arms of each molecular rings have unequal lengths. In 
Fig.~\ref{transmission1} we present the results for such a particular 
configuration where the source and drain are attached to the molecular 
sites $6$ and $9$, respectively. The red, green and blue curves
correspond to the results for the identical parameter values chosen in
Fig.~\ref{transmission}. From the transmission curves (red and green) we 
make out that for a wide energy range across $E=0$, electron conduction does 
not take place, and also the widths of some resonant peaks get reduced 
enormously compared to the symmetric configuration where upper and lower 
arms in each of the two molecular rings are identical 
(Fig.~\ref{transmission}), even though the molecule-to-electrode coupling 
strength is kept unchanged. This is solely due to the effect of quantum 
interference among the electronic waves passing through different arms of 
the molecular rings, and, it can be much clearly analyzed through the 
following arguments. For a fixed molecular coupling, the broadening of 
different resonant peaks which results from the overlap of neighboring
peaks depends on the location of energy levels, as discussed earlier. The 
positions of these energy levels, on the other hand, are directly 
associated with the molecule itself and the real parts of the self-energy 
matrices {\boldmath $\Sigma_S$} and {\boldmath $\Sigma_D$} which correspond 
to the shift of the energy eigenstates of the sample sandwiched between two 
electrodes. Thus for a particular molecule-to-electrode configuration we 
get one set of resonating energy levels, while for the
other configuration a different set of energy levels is obtained. These
generate transmission peaks with different widths associated with the
level spacing. If the molecular coupling strength is low enough, then 
a minor shift of molecular energy levels takes place, and therefore, 
almost identical $T$-$E$ spectrum will be observed for different
molecule-to-electrode configurations. But, for moderate coupling strength
one can regulate electron conduction through the bridge system in a tunable 
way by introducing more asymmetry among these two arms. This behavior is 
nicely reflected in the current-voltage characteristics. The results are 
bestowed in Fig.~\ref{totalcurrent1}. It is clearly observed that for a 
particular bias voltage the current amplitude decreases significantly 
compared to the symmetric configuration, Fig.~\ref{totalcurrent}. The 
symmetry breaking among the molecular arms also tunes the threshold voltage 
$V_{th}$ of electron conduction across the molecular wire for a particular 
twist angle $\theta$, as found by comparing Figs.~\ref{totalcurrent} and 
\ref{totalcurrent1}.

\subsection{Circular current and magnetic field}

Now we focus our attention on the behavior of circular currents and 
associated magnetic fields at the ring centers of the molecule.
These factors are highly sensitive to the molecule itself as well as the
molecule-to-electrode interface geometry. To address these issues, we 
start with the current distribution within the molecule when it is coupled 
to the electrodes in such a way that the upper and lower arms of the 
molecular rings have identical lengths. The current distribution is shown 
in Fig.~\ref{distribution1}, where the blue and green arrows indicate the 
bond currents in upper and lower arms of the rings, respectively. The arrow
sizes represent the magnitudes of bond currents and they are computed when
the bias voltage is fixed at $4$V. Here $I_T$ is the net 
junction current, shown by the red arrow, which is distributed among 
different branches at the junction point. Due to the geometrical symmetry 
reasons, the magnitudes of the bond currents in upper and lower arms of the 
two rings are exactly identical and since they are propagating in opposite 
directions, no net circulating current will appear which results vanishing 
magnetic fields at the ring centers.

In order to establish circular currents in these two molecular rings, we 
\begin{figure}[ht]
{\centering \resizebox*{7.25cm}{4.5cm}
{\includegraphics{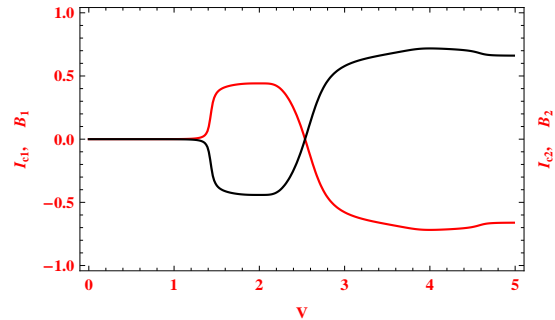}}\par}
\caption{(Color online). Circular currents and associated magnetic fields
at the ring centers of the biphenyl molecule as a function of bias voltage 
when the electrodes are coupled to the molecule following the configuration
prescribed in Fig.~\ref{distribution2}. The red and black curves describe 
the results for the left and right rings, respectively. The twist angle 
$\theta$ is fixed at zero.}
\label{circularcurrent}
\end{figure}
attach the electrodes asymmetrically such that the upper and lower arms of 
the rings have unequal lengths. The internal current distribution for such a 
particular configuration is illustrated in Fig.~\ref{distribution2}, where
we set the same bias voltage as taken in Fig.~\ref{distribution1}. In this
situation the bond currents get unequal magnitudes, and accordingly, circular
\begin{figure}[ht]
{\centering \resizebox*{7.25cm}{4.5cm}
{\includegraphics{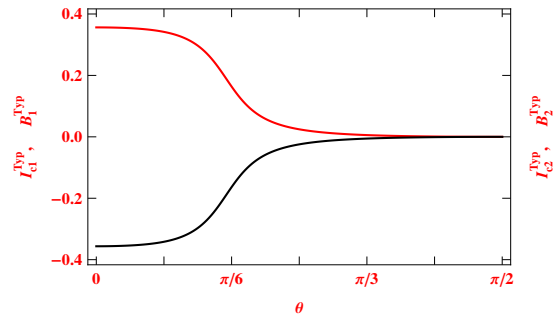}}\par}
\caption{(Color online). Circular currents and corresponding magnetic fields
at the ring centers for a specific voltage bias ($V=1.5$V) as a function of 
twist angle when the molecule is sandwiched between electrodes according to 
the configuration taken in Fig.~\ref{distribution2}. The red and black lines 
correspond to the identical meaning as in Fig.~\ref{circularcurrent}.}
\label{typcircnt}
\end{figure}
currents are established in the two molecular rings. The magnetic fields
at the ring centers associated to these circular currents are also shown. 
In the left ring, the magnetic field is directed along the downward direction, 
represented by encircled cross, while in the other ring it is directed along 
the upward direction, illustrated by encircled dot. We calculate the magnetic 
fields using Biot-Savart's Law, Eq.~\ref{bio}, and scale them in unit of 
$6\mu_0/4\pi R$, where $R$ is the perpendicular distance from the center to
any arm of the molecular ring and the factor $6$ appears due to the 
existence of six bonds in each benzene ring.

Figure~\ref{circularcurrent} demonstrates the magnitudes of the circular 
currents and the associated magnetic fields at the ring centers as a 
function of voltage bias for the planar conformation of the biphenyl 
molecule. From the spectrum we notice that in one voltage regime 
($\sim 0-1.3$V) no circulating current appears, while other voltage regimes 
finite circular currents are available and the sign of these currents also 
changes depending 
on the voltage region. The associated magnetic fields also follow the same 
behavior, as illustrated in the spectrum. This phenomenon can be explained 
as follows. The circular current in a loop geometry is associated with 
energy eigenstates those can be described by current carrying states with 
the current flowing in opposite directions. Now, for a finite bias voltage 
whenever one of these resonant states lies in the Fermi window, associated 
with the applied voltage bias and the nature of voltage drops along the 
molecular wire, we will get the corresponding circular current. When more 
than one resonant states come within this Fermi window, all of them 
contribute to the current which provide a net signal, and, in the particular 
case when they mutually cancel each other, the net signal becomes zero. The
sign of this net circular current or direction of associated magnetic 
field depends on which resonant states dominate the others, which again
certainly depends on the applied bias voltage. From the above analysis we 
can clearly understand the vanishing nature of circulating current in the
above mentioned voltage region ($\sim 0-1.3$V), since up to this voltage
window no resonant state appears which can contribute to the circulating
current. This, on the other hand, is also justified from the red curve 
in $T$-$E$ spectrum, Fig.~\ref{transmission1}, which is drawn for the
planar conformation of the molecule. It indicates that within the energy
window $-0.65$ to $+0.65$ i.e., when the bias voltage becomes $1.3$V,
no electron transmission takes place through the molecule, which results
zero circulating current. For the other voltage regimes finite circular 
currents are available depending on the voltage region.

Finally, we focus on the variation of circular currents and corresponding
magnetic fields at the ring centers as a function of relative twist angle
$\theta$, when the voltage bias is kept constant. The results are shown in
Fig.~\ref{typcircnt}, where two different colored curves represent the 
similar meaning as in Fig.~\ref{circularcurrent}. Quite interestingly we
see that the magnitudes of the magnetic fields at the centers of two 
molecular rings decreases monotonically with relative twist angle $\theta$,
and for large enough $\theta$ they eventually reduce to zero. Thus, for a 
particular bias voltage one can tune the strength of magnetic field 
established at the ring center due to this circular current simply by 
twisting one benzene molecule relative to the other, and hence, by placing 
a local spin or a magnetic ion at the ring center, which will interact to 
the magnetic field, spin selective transmission can be achieved through 
this molecular system. This conformation-dependent spin selective 
transmission will be investigated in a recent forthcoming paper. 

\section{Conclusion}

In conclusion, we have investigated in detail the conformation-dependent 
two-terminal electron transport through a biphenyl molecule within a simple 
tight-binding framework using Green's function formalism. Two principal 
results have been obtained and analyzed. First, the dependence of molecular
twist on electronic transmission probability and the overall junction 
current have been discussed. Our results lead to a possibility of getting the 
conformation-dependent switching action using this molecule. Second, we
have investigated the variation of circular currents and associated 
magnetic fields developed at the ring centers as a function of the
relative twist angle. Tuning this angle, one can tailor the strength of 
magnetic field at the ring centers, and, we believe that the present 
analysis may provide the possibilities to design molecular spintronic 
devices using organic molecules with loop substructures.

Throughout our work we have ignored the inter- and intra-site Coulomb 
interactions as well as the effect of the electrodes, which we plan
to consider in our future works. Another important assumption is the 
zero temperature approximation. Though all the results presented in this 
communication are worked out at absolute zero temperature limit, the 
results should remain valid even at finite temperatures ($\sim300\,$K) 
since the broadening of the energy levels of the biphenyl molecule due 
to its coupling with the metal electrodes is much higher than that of the 
thermal broadening~\cite{datta}.

\section{Acknowledgment} 

The author is thankful to Prof. Abraham Nitzan for many stimulating 
discussions.

\end{document}